\documentclass[prd,aps,twocolumn,nofootinbib,preprintnumbers]{revtex4}
\usepackage{amsmath,amssymb,epsf}
\def\be{\begin{equation}}
\def\ee{\end{equation}}
\def\ba{\begin{equation}}
\def\ba{\begin{eqnarray}}
\def\ea{\end{eqnarray}}
\def\lsim{~\mbox{\raisebox{-.6ex}{$\stackrel{<}{\sim}$}}~}
\def\gsim{~\mbox{\raisebox{-.6ex}{$\stackrel{>}{\sim}$}}~}
\def\bq{\begin{quote}}
\def\eq{\end{quote}}

\def\lmk{\left(}
\def\rmk{\right)}

\begin{document}
\title{Equation of state and Beginning of  Thermalization After Preheating}
\author{Dmitry Podolsky$^{1}$\footnote{On leave from Landau Institute
for Theoretical Physics, 117940, Moscow, Russia}, Gary
N. Felder$^{2}$, Lev Kofman$^{1}$ and Marco Peloso$^{3}$}
\affiliation{${}^1$CITA, University of Toronto, 60 St. George st.,
Toronto, ON M5S 3H8, Canada}
\affiliation{${}^2$Department of Physics, Clark Science Center, Smith College 
Northampton, MA 01063, USA}
\affiliation{${}^3$School of Physics and Astronomy, University of 
Minnesota, Minneapolis, MN 55455, USA}
\date{\today}

\begin{abstract}

We study the out-of-equilibrium nonlinear dynamics of fields after
post-inflationary preheating. During preheating, the energy in the
homogeneous inflaton is exponentially rapidly transfered into highly
occupied out-of-equilibrium inhomogeneous modes, which subsequently
evolve towards equilibrium. The infrared modes excited during
preheating evolve towards a saturated distribution long before
thermalization completes. We compute the equation of state during and
immediately after preheating. It rapidly evolves towards radiation
domination long before the actual thermal equilibrium is
established. The exact time of this transition is a non-monotonic
function of the coupling between the inflaton and the decay products,
and it varies only very weakly (around $10^{-35}$ s) as this coupling
changes over several orders of magnitude. This result is applied to
refine the relation between the number of efoldings N and the physical
wavelength of perturbations generated during inflation. We also
discuss the implications for the theory of modulated perturbations
from preheating. We finally argue that many questions of the thermal
history of the universe should be addressed in terms of
pre-thermalization, illustrating this point with a calculation of
perturbative production of gravitinos immediately after chaotic
inflation.  We also highlight the effects of three-legs inflaton
interactions on the dynamics of preheating and thermalization in an
expanding universe.

\vskip 1cm

\end{abstract}

\preprint{UMN-TH-2407/05}

\maketitle
\section{Introduction: Between Inflation and Thermalization}\label{introduction}
According to the inflationary scenario, the universe at early times
expands quasi-exponentially in a vacuum-like state without entropy or
particles. During this stage of inflation, all energy is contained in
a classical slowly moving inflaton field $\phi$.  The fundamental
Lagrangian ${\cal L}(\phi, \chi, \psi, A_i, h_{ik}, ...)$ contains the
inflaton part with the potential $V(\phi)$ and other fields which give
subdominant contributions to gravity. In chaotic inflationary models,
soon after the end of inflation the almost homogeneous inflaton field
$\phi(t)$ coherently oscillates with a very large amplitude of the
order of the Planck mass around the minimum of its potential.  Due to
the interactions of other fields with the inflaton in ${\cal L}$, the
inflaton field decays and transfers all of its energy to relativistic
particles.  If the creation of particles is sufficiently slow (for
instance, if the inflaton is coupled only gravitationally to the
matter fields) the decay products simultaneously interact with each
other and come to a state of thermal equilibrium at the reheating
temperature $T_r$. This gradual reheating can be treated with the
perturbative theory of particle creation and thermalization as long as the couplings are sufficiently small
\cite{DL,therma,GKR}.  However, the particle production from a
coherently oscillating inflaton for a wide range of couplings occurs in
a non-perturbative regime of parametric
excitation~\cite{KLS94,TB,KLS97}.  This picture, with variation in its
details, is extended to other inflationary models. For instance, in
hybrid inflation inflaton decay proceeds via a tachyonic instability
of the inhomogeneous modes which accompany the symmetry breaking
\cite{tach}.

One consistent feature of preheating -- non-perturbative copius
particle production immediately after inflation -- is that the process
occurs far away from thermal equilibrium.  The energy of the inflaton
zero mode is transferred to particles in an out-of-equilibrium state
with very large occupation numbers within a very short time interval
of about $10^{-35}$ sec, which can be much shorter than the time
needed for relaxation towards thermal equilibrium.

One can expect the initial conditions for many parameters of the
subsequent cosmological thermal history to be settled during or around
preheating.  It is often thought that the details of the transition
between inflation and the hot radiation dominant stage are not
relevant, except for the so-called reheating temperature $T_r$, the
temperature of the ultra-relativistic plasma at the time when it
reaches thermal equilibrium. Definitely, this is an important
parameter of the early universe.  However, a precise understanding
of how thermal equilibrium is reached is crucial since partial thermal
distributions can be responsible for cosmological baryo/leptogenesis,
the possible creation of dangreous cosmological relics, etc.  The
out-of-equilibrium character of preheating opens the possibility for
equally relevant phenomena associated with non-equilibrium physics,
including phase transitions, non-thermal production of heavy
particles, etc.

Precise knowledge of the expansion of the universe $a \left( t
\right)$ is also required to connect a physical wavelength of the
observed cosmological perturbation, $k / a \left( t \right)$, to the
number of e-foldings $N$ at which that wavelength exited the horizon
during inflation.

Finally, we can have an alternative mechanism of generation of
(almost) scale free adiabatic metric perturbations from preheating,
based on spatial modulation of couplings~\cite{mod,mod1,BKU}.  Small
spatial fluctuations in the couplings lead to small fluctuations of
the rates of physical processes, which in turn generate adiabatic
perturbations of the energy density and of the metric. Analyzing the
details of this mechanism require precise knowledge of the evolution
of the equation of state (EOS).

To understand the early post-inflationary period we have to understand
the dynamics of interacting fields in an out-of-equilibrium state with
large occupation numbers evolving towards a state of ultimate thermal
equilibrium.  This is a complicated problem of non-equilibrium
quantum field theory, which is by itself a very interesting topic.

The theory of the transition from inflation to thermalization has been
investigated with fully non-linear numerical lattice simulations plus
different techniques of classical field dynamics
\cite{KT,FK,MT04,FT}. Classical field dynamics is adequate as long as
occupation numbers of field excitations are large.  There has also
been progress in understanding out-of-equilibrium QFT dynamics in
$O(N)$ sigma models beyond the Hartree approximation
\cite{Boyanovsky:1997cr}, taking into account crucial effects of
rescattering after preheating \cite{J}. From all of these methods, the
following picture emerges. Immediately after preheating, either from
parametric resonance or tachyonic, one or more bose fields are excited
in an out-of-equlibrium state with large occupation
number. Backreaction of these fields terminates their
production. Interaction (rescattering) of these modes between each
other, and with the remaining inflaton field after the first stage of
preheating is violent and non-perturbative. During this very short
stage a large amount of entropy is generated, chaotic (turbulent) wave
dynamics is established, and the fields not directly excited during
preheating get excited in out-of-equilibrium states due to
rescattering; residual inflaton oscillations are still present. A
next, longer stage then takes place, characterized by lower (but still
large) occupation numbers such that rescattering is perturbative, and
the occupation spectra gradually move towards a saturated state by
cascading towards ultra-violet (UV) and infra-red (IR) modes (in the
spirit of Kolmogorov wave turbulence). The last and the longest stage
will be the stage of proper thermalization, when the distributions
evolve towards thermal equilibrium.  Quantum physics is important at
the end of this stage, when the occupation numbers are small and the
classical approximation breaks down.  One may say that reheating is
completed when this last, longest stage is completed, and its timing
defines $T_r$ and $a_r$.

In this paper we suggest a new look at the transition between
inflation and thermalization.  We focus on calculation and
understanding of the effective equation of state $w=p/\epsilon$
throughout all of these stages after inflation.  We use numerical
lattice simulations \cite{FT} to calculate $w$ in the simple chaotic
model with a quadratic potential and a simple four-legs interaction
$g^2 \phi^2 \chi^2$.

In particular, we notice that $w$ approaches (but does not necessarily
reach) the equilibrium radiation-dominated value $1/3$ while the
fields are far from ultimate thermal equilibrium. We compare this with
the analysis of~\cite{Berges:2004}, devoted to thermalization of the
${\rm O} \left( N \right)$ sigma model in Minkowski spacetime. In that
case, the equation of state was found to evolve sharply towards
$w=1/3$ long before thermalization completes, which prompted the
authors of~\cite{Berges:2004} to describe the state of the system as a
pre-thermalized state. Although we also see a somewhat similar effect,
the expansion of the universe and the presence of a residual massive
inflaton component, prevent the equation of state from being exactly
the one of radiation at this very early stage. Moreover, we enphasize
that the exact microphysics of the system is described by a turbulent
state, which is still very far from a thermalized one. (This difference
can be important when discussing the cosmological effects of decay
products, as for instance production of dangerous relics from
nonthermal distributions.) Indeed, our simulations show that the
occupation numbers of excited infra-red modes evolve towards a
steady state, related to turbulence of classical interacting waves.

The plan of the paper is the following:

In Section \ref{sec:model} we describe preheating in a simple chaotic
inflationary model. In Section \ref{sec:numerics} we present the
results of lattice simulations for time-dependent variables, including
the equation of state $w(t)$ during different stages of preheating,
occupation numbers and fluctuations.

In Section \ref{sec:implementations} we discuss some immediate
applications of our results for cosmology.  We consider the
application of the EOS evolution to modulated cosmological
fluctuations and to the $N-\log k$ formula. We note several
qualitatively important issues in the dynamics of thermalization.  The
interaction $\phi^2 \, \chi^2$ describes (at the perturbative level)
scattering between inflaton quanta, rather than inflaton decay.
Scattering between massive $\phi$ soon becomes inefficient, and does not
lead to a complete depletion of the inflaton, which eventually ends up
dominating over the light degrees of freedom $\chi$.  This problem is
automatically avoided if three-legs bosonic interactions
$\phi \, \chi^2$ are also present.  We illustrate this with an example
motivated by supersymmetry. We also discuss the rapid saturation of
the IR modes excited during preheating, which occurs long before the
ultimate thermal equilibrium (of all modes) is reached.

One consequence of the excitation of infrared modes is that it can
lead to an overproduction of gravitinos or of other dangerous
gravitational relics. This generation is different from direct
nonperturbative production during preheating~\cite{grapreh}. Instead,
it is analogous to the perturbative thermal production first
considered in~\cite{grath}. The main difference with~\cite{grath} is
that we compute the production right after the first preheating stage,
and not only after thermalization has completed~\cite{grath}. Although
the distributions formed at preheating/rescattering are far from
thermal, the high particle occupation numbers can lead to a
significant generation of gravitinos.

The above results are summarized in Section \ref{sec:summary}.

\section{The models and methods}\label{sec:model}

In this section we briefly describe the model of chaotic inflation
with a quadratic potential and additional interaction terms and the
methods we use to investigate the fields dynamics. The quadratic form
of the potential is a good generic approximation around the
minimum. For simplicity, we take the potential
\be
\label{pot}
V(\phi)= \frac{1}{2} m^2 \phi^2
\ee
also throughout inflation. The inflaton mass is taken to be
$m\simeq 10^{13} \,$ Gev, which, in the simplest cases,
reproduces the correct amplitude for the CMB anisotropies.
After inflation, the background inflaton
field oscillates with its amplitude diluted by the expansion as $\phi
\propto \phi_0 / a^{3/2} \,$, where $\phi_0 \approx M_p/10$ is the
initial amplitude, and $a(t)$ is the scale factor of the FRW flat
universe. More accurately, the inflaton evolution is given by $\phi
(t) \approx \frac{M_P}{\sqrt{3\pi}mt} \sin mt$.  The energy density of
the inflaton oscillations evolves as $\epsilon \approx m^2
\phi_0^2/a^3$, while the pressure is $p \approx -(m^2 \phi_0^2/a^3)
\cos 2mt$. Since the frequency of the oscillations is higher than the
rate of expansion $H$, we deal with a nearly vanishing averaged
effective pressure, $p \approx 0$. Therefore, immediately after
inflation, the equation of state (EOS) is effectively the one of
matter domination
\be
\label{eos1}
w=p/\epsilon \approx 0 \ .
\ee
This is what we expect for inflaton oscillations interpreted as the
coherent superposition of heavy inflaton (quasi)-particles at rest.

Inflaton oscillations decay due to the coupling between the inflaton
and other fields.  One may think about the inflaton coupling to other
bosons through four-legs or three-legs interactions, for instance
coupling to another scalar $\chi$ as $g^2\phi^2\chi^2$ or $g^2 \sigma
\phi \chi^2$. Yukawa-type couplings to fermions will be of the form $h
\bar \psi \phi \psi$. During preheating the leading channel of
inflaton decay will be parametric resonance decay into bosons. A three
legs interaction between bosons $g^2 \sigma \phi \chi^2$ requires a
dimensionful scale $\sigma$, which emerges naturally in SUSY theories. 
However, in most cases the amplitude of the inflaton is sufficiently
high during the early stages of preheating that the four-legs interaction dominates \cite{new}.
For this reason, at this stage we consider only the four-legs interaction
\be\label{int}
{\cal L}_{int}=- \frac{g^2}{2} \phi^2 \chi^2 \ .
\ee
Three-legs interactions are important for the completion of the decay
of massive inflatons, and we will discuss them further in Section
\ref{sec:implementations}, and in detail in \cite{new}.

Before proceeding with the generation of $\chi$ particles from the
inflaton oscillations, we need to comment on the range of the coupling
$g^2$.  At the level of QFT we can obtain an upper limit for $g^2$.
Indeed, in non supersymmetric models, the interaction (\ref{int})
leads to radiative corrections~\footnote{There is also a divergent
quadratic contribution. However, for the supersymmetric theory this
term is independent of the value of $\phi \,$ and can thus be
cancelled by a counter-term.} to the effective potential (\ref{pot}),
of the form $\frac{g^4 \phi^4}{32\pi^2} \ln \phi$.  For inflaton
values $\phi \sim 4 M_p \,$, which correspond to the scales where
cosmological fluctuations are observed, radiative corrections do not
alter the the potential (\ref{pot}) as long as $g^2 \lsim 10^{-5}$. In
supersymmetric theories radiative corrections from bosons and fermions
cancel each other out so that the value of $g$ can be even higher.

There is also a lower limit on $g^2$.  During inflation the field
$\chi$ has the effective mass $m_\chi^2 + g^2 \langle \phi^2 \rangle
\sim g^2 M_p^2 \,$, where $m_\chi$ is its ``bare mass'': we take
$m_\chi \ll m \,$ and thus neglect it. It is known \cite{KL87} that if
we have two scalar fields $\phi$ and $\chi$, the latest stage of
inflation will be driven by the lightest scalar (light scalars are
always generated during inflation). A consistent setting where the end
of inflation is driven by the $\phi$ field requires $ g^2 M_p^2 >
m^2$, i.e. $g^2 > 10^{-12}$. This lower limit can be evaded, however, if
$\chi$ is conformally coupled to gravity.

A quantum scalar field $\chi$ in a flat FRW background has the
eigenfunctions $\chi_k(t) e^{-i \vec k \vec x}$ with comoving momenta
$\vec k$.  The temporal part of the eigenfunction $\chi_k(t)=a^{3/2}
X_k(t)$ obeys the equation
\be \ddot X_k + \lmk
\frac{k^2}{a^2} + g^2 \phi^2 \rmk X_k = 0
\label{modesofchi}
\ee
with positive-frequency initial conditions in the far past.  In
the effective frequency $\omega_k^2 \approx \frac{k^2}{a^2} + g^2 \phi^2$ we
neglect small terms proportional to $H^2$.  This is an oscillator-like
equation with a periodic frequency, which contains exponentially
unstable solutions corresponding to parametric resonance. It is
convenient to use a new time variable $z=mt$.  Then, the essential
dimensionless coupling parameter which characterizes parametric
amplification is $q=\frac{g^2 \phi_0^2}{4 \, m^2}$.  
The limits on the coupling $g^2$ we have discussed above
correspond to $10^{-3} \lsim q \lsim 10^5$.  A large portion of this
interval is in the broad parametric resonance regime ($q > 1$) of
production of $\chi$ particles~\cite{KLS94}.

  What is important, however, is that even one
bosonic field with a large parameter $q$ is sufficient to have the
leading channel of inflaton decay occur through broad parametric
resonance. Indeed, if one such field $\chi$ is exponentially
amplified in this way then other fields that are coupled to it
directly or indirectly will tend to be amplified as well \cite{FK}.

We are interested in the dynamics of the fields $\phi$ and $\chi$
after inflation. Although this toy model is simple, its dynamics are
rather rich and complex. The first stage is the stage of
parametric excitation of the quantum fluctuations of the field $\chi$,
where the light $\chi$ particles are copiously produced from the
classical inflaton oscillation.  This process is described by the
theory of broad stochastic parametric resonance in an expanding
universe \cite{KLS97}.

As soon as the occupation number of created particles
\be
\label{particles}
n_k = \frac{\omega_k}{2} \lmk \frac{|\dot X_k|^2}{\omega_k^2} + |X_k|^2 \rmk 
- \frac{1}{2}
\ee
becomes large, $n_k \gg 1$, one can approximate the dynamics with that
of the classical interacting fields $\phi$ and $\chi$.  Fully
non-linear wave equations for coupling classical fields are rather
complicated.  We use the LATTICEASY numerical simulations
\cite{FT} to solve the equations $\Box \phi +m^2\phi +g^2\phi \chi^2 = 0$
and $\Box \chi +g^2 \phi^2 \chi = 0$ in an expanding universe ($\Box$
being the covariant wave operator in the FRW geometry).  The scale
factor $a(t)$ is self-consistently calculated from the Friedmann
equation.  The initial conditions correspond to gaussian initial
fluctuations of $\phi$ and $\chi$, which arise from vacuum fluctuations.

The novel element of our current study is the focus on new diagnostics
of the numerical experiments.  In particular, we investigate the
evolution of the equation of state $w(t)$ during different stages, and
for various couplings $g^2$.  Many previous numerical studies of
preheating concentrated on inflation with a quartic potential
$V(\phi)=\lambda \phi^4$, where the expansion of the universe can be
``scaled away''. The inflaton oscillations in this model have the EOS
$w=1/3$, which corresponds to that of radiation domination, so $w$ does not
change after the inflaton decay. By contrast, $w$ changes during
preheating in the more general model with a quadratic potential.

The total energy density $\epsilon$ and pressure $p$ contain
contributions from kinetic and gradient energies of the fields $\phi$
and $\chi$, plus interaction terms. It is instructive to calculate
separate contributions from all of these terms. We find that the
interaction term, except for a transient moment between the end of
preheating and onset of turbulent regime, is sufficiently small that
we can interpret the results in terms of different particles $\delta
\phi$ and $\chi$ coupled to each other. In particular, this allows us
to interpret the energy density in each component in terms of the
integral
\be
\label{energy}
\epsilon\approx \frac{1}{(2\pi)^3 a^4} \int d^3k \, \omega_k \, n_k \ ,
\ee
where the comoving frequency is 
\begin{equation}
\omega_k= \sqrt{k^2+ m_{eff}^2 a^2} \,\,,
\label{comom}
\end{equation}
and the co-moving occupation number $n_k$ is given in
(\ref{particles}). To study $w$ we have to monitor the relationship
between the effective masses of the fields
\be
\label{mass1}
m_{\phi,eff}^2=m^2 +g^2 \langle \chi^2\rangle \ ,
\ee
\be
\label{mass2}
m_{\chi,eff}^2=g^2 \langle \phi\rangle^2 +g^2 \langle  {\delta \phi}^2\rangle \ .
\ee
and the typical momenta.

The variances of the fluctuations
\be
\label{fluc1}
 \langle \chi^2\rangle=\frac{1}{(2\pi)^3 } \int d^3k \, |\chi_k|^2
\ee
\be
\label{fluc2}
 \langle  {\delta \phi}^2\rangle =\frac{1}{(2\pi)^3 } \int d^3k \, |\delta \phi_k|^2
\ee
are also an important diagnostic of the non-equilibrium physics.  Let us
rewrite the wave function $X_k(t)$ in the WKB form
$X_k(t)=\frac{\alpha_k}{\sqrt{2 \omega_k}}\, e^{-i\int dt \omega_k}
+\frac{\beta_k}{\sqrt{2 \omega_k}}\, e^{i\int dt \omega_k}$.  While
the occupation numbers in terms of $\alpha-\beta$ coefficients are
simply $n_k=|\beta_k|^2$, the exact expression for the variances is
more complicated
\be
\label{fluc3}
 \langle \chi^2\rangle=\frac{1}{(2\pi)^3 a^3} \int \frac{d^3k}{\omega_k} \,
 \left(n_k+ 2 Re \left( \alpha_k \beta_k^* e^{2 \int dt i\omega_k} \right) \right) \ .
\ee
Particles $\chi$ produced at preheating are created in a squeezed
state.  In particular, this is reflected in the presence of the
oscillatory second term in the integral (\ref{fluc3}), where the
phases of $\alpha_k ,\, \beta_k$ are present.

The total co-moving number densities of $\chi$ and $\phi$ particles
are \be
\label{part1}
 N_{\chi}= \frac{1}{(2\pi)^3 a^3} \int d^3k \, n^{\chi}_k \ ,
\ee
\be
\label{part2}
 N_{\phi}= \frac{1}{(2\pi)^3 a^3} \int d^3k \, n^{\phi}_k \ ,
\ee
respectively.  It is instructive to monitor the time evolution of the
total number of particles $N_{\rm tot}=N_{\chi}+N_{\phi}$.  At leading
order in $g^2$ the total number of particles is conserved, $N_{\rm
tot} \left( t \right) = {\rm const}$. However, there is a brief period
immediately after preheating where $N_{\rm tot}$ is poorly defined due
to the strong interactions between the waves. Indeed, during
preheating the occupation number of $\chi$ particles can be as large
as $n_k \sim 1/g^2$. Consider the collision integral in the kinetic
equation for the particle. If we formally represent it as a series
with respect to the coupling $g^2$, the smallness of $g^2$ will be
compensated by the large occupation numbers.  This means that the next
terms of the collision integral associated with the higher order
diagrams (for example, the diagram with four incoming, two outgoing
legs, and two vertices) are as important as the first term.  In fact,
the non-perturbative character of the interactions during this short
intermediate stage makes it difficult to investigate the whole
dynamics analytically. After this period ends, the spectrum of the
occupation numbers formed at preheating is weighted towards the
infra-red relative to a thermal distribution of the same energy
density.  Hence, particle fusion is more favored than particle
dissociation. Once rescattering leads to a decrease of $n_k$, only the
leading four-legs diagram with one vertex becomes important, and
$N_{\rm tot}$ decreases on the way to thermalization, on a timescale
which is much longer than the one which can be observed in our lattice
simulations.

\section{Output of the Calculations}\label{sec:numerics}

\subsection{The Calculations}

We performed three-dimensional lattice simulations for the model of
Section \ref{sec:model}. Our grid was a $256 \times 256 \times 256$
cube with a comoving edge size $L=10/m$, which corresponds to a
comoving grid spacing of $dx \approx 0.04/m$. As energy flows towards
the UV end of the spectrum the simulations eventually reach a point
where the grid spacing is too large to capture the important UV
physics. By monitoring the spectra of the fields, however, we can
verify that these simulation parameters were adequate to capture the
relevant IR and UV physics well past the end of preheating. The time
step was $dt=0.001/m$ and the inflaton mass was $m=10^{-6} M_p$. We
used values of the coupling near $g^2 = 10^{-7}$ This value is optimal
because it is large enough to produce highly efficient preheating, but
small enough that the occupation numbers $n_k \sim 1/g^2$ produce
strong rescattering. The results should be qualitatively similar for a
wide range of values of $g^2$, but would require more IR and/or more
UV to simulate accurately.

To probe later times and wider ranges of the couplings it will be
necessary to extend the lattice simulations. This can be done with a
parallelized version of the simulation code LATTICEEASY (currently
under construction), or by combining the straightforward lattice
simulations with other methods, like the equations for a large number
of weakly coupled oscillators \cite{ST}. We intend to pursue both of these
approaches in subequent work.

In the rest of this section we present the results of our
simulations.

\subsection{Equation of State}

\begin{figure}[h]
\leavevmode\epsfxsize=\columnwidth \epsfbox{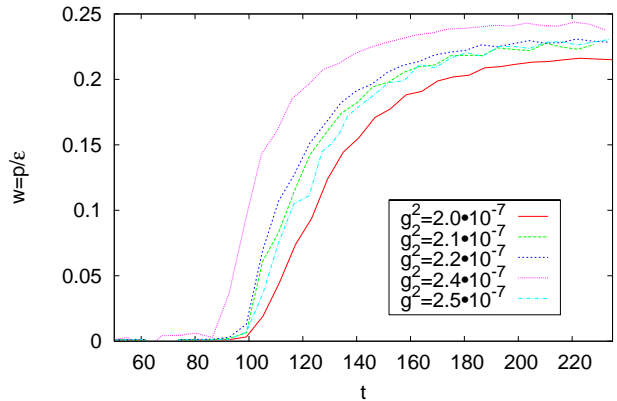}
\caption{Evolution of the equation of state  $w = w(t)$ as a function of time (given in units of $m^{-1}$)
for various couplings $g^2$ around $g^2 = 2 \times 10^{-7}$.}
\label{fig:w1}
\end{figure}

The time evolution of the EOS $w(t)$ for different couplings is shown in
Figure~\ref{fig:w1}. Each
point plotted on this figure represents the value of $w$ averaged over
a complete inflaton oscillation.
This represents one of the main results of our study.

Immediately after inflation, the EOS averaged over inflaton oscillations is $w=0$.
It sharply changes at the end of preheating.

There are at least three important points worth enphasizing about the
evolution of $w$.

i) {\it First, the transition of the EOS from $w=0$ to the value $w
\sim 0.2 - 0.3$ occurs very sharply, within a time interval $\sim
10^{-36}$ sec.}

Indeed, recall that the unit of time on the plots is $1/m$, where
$m$ is the inflaton mass, i.e. $10^{-37}$ sec. The first stage of
preheating is completed within about a hundred of these units, i.e.,
$10^{-35}$ sec. The rise of $w$ and gradual saturation takes roughly
the same time.

ii) {\it Second, the dependence of $w(t)$ on the coupling $g^2$ for
resonant preheating is a non-monotonic function of $g^2$.}

\begin{figure}[h]
\leavevmode\epsfxsize=\columnwidth \epsfbox{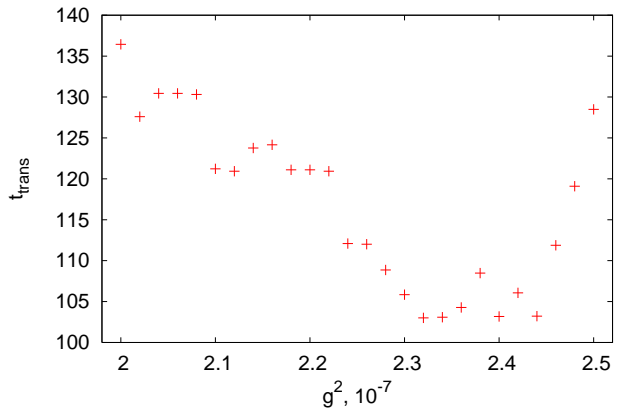}
\caption{Transition (preheating)  time as a function of $g^2$.}
\label{fig:tran}
\end{figure}

This is to say that the time during which preheating comes to an end
is very weakly (logarithmically) dependent on the coupling. As seen
from Figure~\ref{fig:w1} the curves $w(t)$ begin to shift to the left
towards an earlier end of preheating, as we vary $g^2$ by $5 \%$.
However, at some point the curves stop moving to the left and instead
begin to return toward the right. As we change $g^2$ by about $25 \%$,
the cycle repeats.  As we vary $g^2$, the function $w$ not only
shifts, but it also varies its detailed shape.  Still, to characterize
these variations, we pick up the moment where $w$ is equal to the
value $0.15$ (just for convenience of calculation),
$w(t_{tran})=0.15$. This allows us to plot the transition moment
$t_{tran}(g^2)$ as a function of $g^2$, see Figure~\ref{fig:tran}.

We see that the transition time varies between $100/m$ and
$150/m$.  This non-monotonic behavior of the duration of preheating
is explained in the theory of broad paremetric resonance \cite{KLS97}
(see Sections 6 and 9 there).

The $g^2$ dependence of the EOS is the critical issue for the theory
of modulated cosmological perturbations, which we will discuss in
Section \ref{sec:implementations}.

iii) {\it The third point is that $w$ does not necessarily immediately
go to the radiation dominated value $1/3$.} 
This is partly because immediately after preheating 
the light field still has a significant induced effective mass due to
the interaction, and partly due to the significant residual contribution from
the homogeneous inflaton \cite{MT04}.
 Unfortunately, limitations on running longer simulations
preclude us from seeing further details of the time evolution of $w$.
However, we have a strong theoretical argument to advance the discussion further.
In a model with a massive
inflaton and light scalar $\chi$ even the radiation dominated stage is
transient. Indeed, sooner or later the massive inflaton particles,
even if significantly under-abundant at the end of preheating, will
become the dominant component, and the universe will again be
matter-dominated.

\subsection{Occupation Numbers}

\begin{figure}[h]
\leavevmode\epsfxsize=\columnwidth \epsfbox{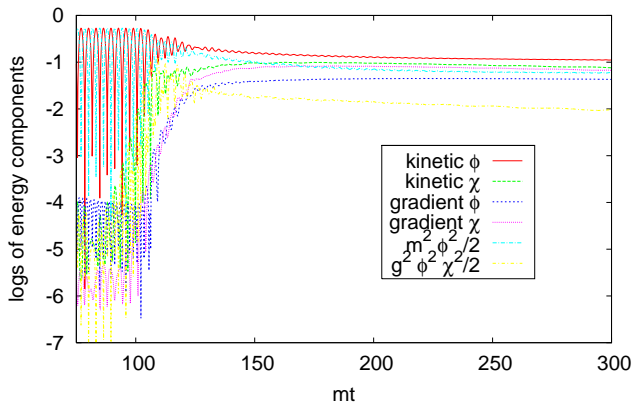}
\caption{Relative contribution of each of the energy components to the
total energy, as a function of time.  The vertical axis is the $\log$ of
the various energy components in units of the initial energy $m^2\phi_0^2$
multiplied by $a^3$.}
\label{fig:comp}
\end{figure}

The occupation numbers $n^{\chi}_k$ and $n^{\phi}_k$ are among the most
interesting variables to understand the micro-physics in our system of
two interacting fields. First, we shall determine when  the definition of $n_k$ is
meaningful.  To do so, we consider the composition of total energy
density $\epsilon$ of the system of coupled $\phi$ and $\chi$
fields.  The total energy density $\epsilon$ can be decomposed into partial contributions from
the kinetic energy of both fields, their gradient energy, their
potential energy ($\phi$ only in this model), and finally the interaction energy
\be
\label{decomp}
\epsilon=\frac{1}{2}\dot \phi^2+\frac{1}{2}\dot \chi^2+\frac{1}{2a^2}(\nabla \phi)^2+
\frac{1}{2a^2}(\nabla \chi)^2+\frac{1}{2} m^2 \phi^2+\frac{1}{2} g^2 \phi^2 \chi^2 \ .
\ee
Figure \ref{fig:comp} shows the relative contribution to the total
energy from each of the components for $g^2=2.5 \times 10^{-7}$.

We note two features of this plot. First, the interaction term is
comparable to the other terms in the time interval $100/m < t <
120/m$. In this short period, the formula (\ref{energy}) for the
energy of the particles is not a good approximation and the occupation
number $n_k$ is not well defined. Outside this time
interval, $n_k$ is a meaningful quantity. Secondly, the contribution
from the background homogeneous inflaton is dominant even after
preheating, up to approximately $t \sim 150/m$. A similar point was
made for the $\lambda \phi^4$ model in ~\cite{MT04}.

Let us now turn to the occupation numbers $n_k$. We find it more
instructive to output not the occupation numbers $n_k$ {\it per se}
(as it is commonly done in the literature) but the combination of
$n_k$ with the energy per mode $\omega_k \,$. This combination can be
immediately compared with the Rayleigh-Jeans spectrum,
\be
\label{thermal}
n_k \approx \frac{T_{eff}}{\omega_k -\mu} \ ,
\ee
which corresponds to the equipartition spectrum of classical waves (we
introduce the chemical potential $\mu$ for generality). The comparison
allows to determine how close the distribution is to the thermal one.

\begin{figure}[h]
\leavevmode\epsfxsize=\columnwidth \epsfbox{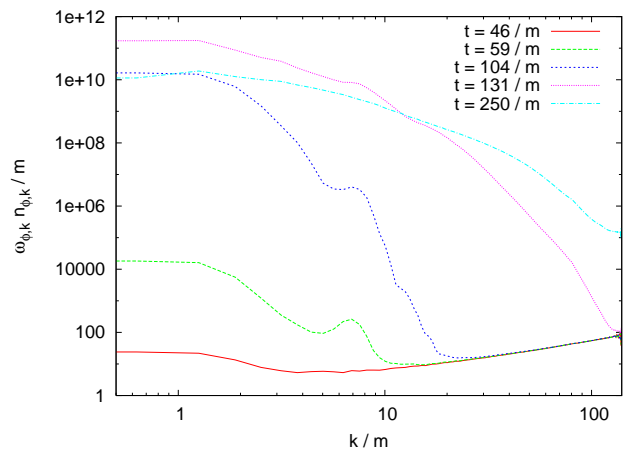}
\caption{Time evolution of the combination $\omega_{\phi, k}
n_k^{\phi}$, for the model $g^2=2.5\cdot 10^{-7}$.}
\label{fig:phinkwk}
\end{figure}

\begin{figure}[h]
\leavevmode\epsfxsize=\columnwidth \epsfbox{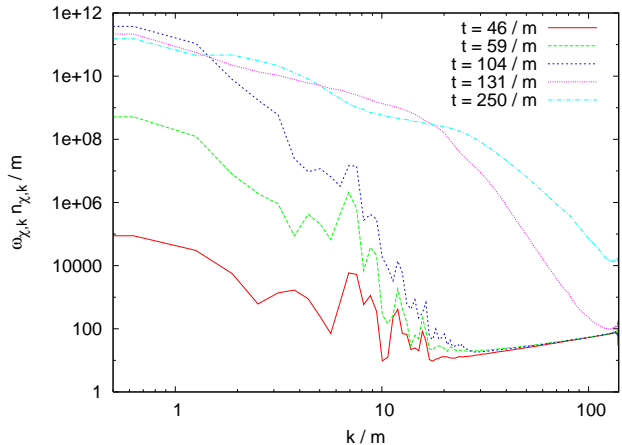}
\caption{Same as Fig.~\ref{fig:phinkwk}, but for the $\chi$ field.}
\label{fig:chinkwk}
\end{figure}

The combination $n_k \, \omega_k$ for the two fields is shown at some
characteristic times in the two Figures \ref{fig:phinkwk} and
\ref{fig:chinkwk}. There are three distinct stages which characterize
the evolution covered by our simulation. The first stage,
characterized by linear dynamics, is the one of preheating and early
rescattering. The first modes to be populated are the IR
ones. Preheating of $\chi$ particles occurs in the resonant band at
comoving momentum $k_* = \sqrt{g \, m \, \phi_0} \, a^{1/4} \simeq 7
\, m \, a^{1/4}$~\cite{KLS97}. Then, quanta $\delta \, \phi$ are
generated by rescattering. The annihilation $\delta \chi_k \, \delta
\chi_k \rightarrow \delta \phi_k \, \delta \phi_k$ amplifies quanta of
the inflaton at $k \simeq k_* \,$. Even more effective is the
rescattering of the $\chi$ quanta against the inflaton zero mode,
$\delta \chi_k \, \phi_0 \rightarrow \delta \chi_k \, \delta \phi_k
\,$, which produces quanta of both $\phi$ and $\chi$ at momentum $k
\simeq k_* / 2 \,$.

The second stage is a violent stage of highly nonlinear
dynamics. Starting at $m t \sim 100-110 \,$, the higher band at $k_*
/2$ both increases in its amplitude and broadens towards higher
momenta; quite interestingly, the peak location shifts from $\sim
k_*/2$ to $\sim k_*$ during this quick and explosive stage. As we
remarked, the particle occupation number is ill defined at this stage,
and a nonlinear wave description is more adequate.

The explosive stage of rescattering ends at about $m t \sim 130
\,$. In the next stage, characterized by perturbative dynamics, the
distributions smooth out and start evolving (at a much slower rate)
towards higher comoving momenta. The spectra in the IR approach a
saturated power-law state, which then slowly propagates towards the
UV.  Although one can observe a greater tendency towards
thermalization for the IR modes (where the rescaled spectra are closer
to flat), the overall distributions are still typical of the turbulent
regime, and they are far from thermal. If we contrast this with the
macroscopic behavior we have described above, particularly the
evolution of the EOS, we see that the system can be considered in a
pre--thermalized state (see also \cite{Berges:2004}), but that
thermalization is still far from being complete .

It is also instructive to consider the product of the occupation
number $n_k$ with the phase space sphere area $k^2$ and energy per
mode $\omega_k$. This combination represents the energy density of the
quanta at momentum $k$ (since the product $4 \, \pi \, k^2 \, \omega_k
\, n_k \, d k$ is the energy density of the quanta whose momentum has
a magnitude between $k$ and $k+d k \,$). 

\begin{figure}[h]
\leavevmode\epsfxsize=\columnwidth \epsfbox{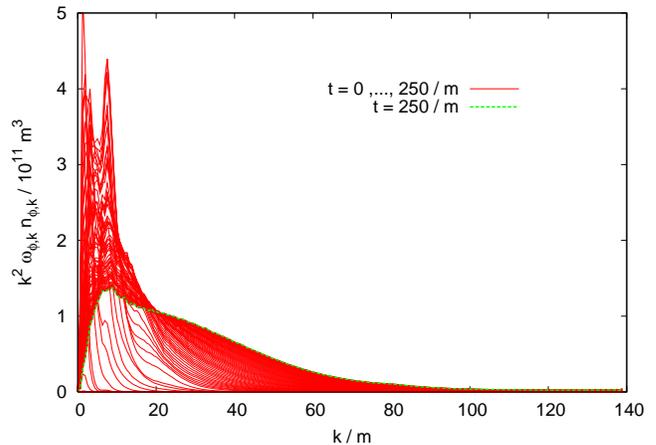}
\caption{Time evolution of the combination $k^2 \omega_k n_k$ for the
field $\phi \,$. The thicker (green) curve is the spectrum at the
final time of our evolution.}
\label{fig:phispL}
\end{figure}

\begin{figure}[h]
\leavevmode\epsfxsize=\columnwidth \epsfbox{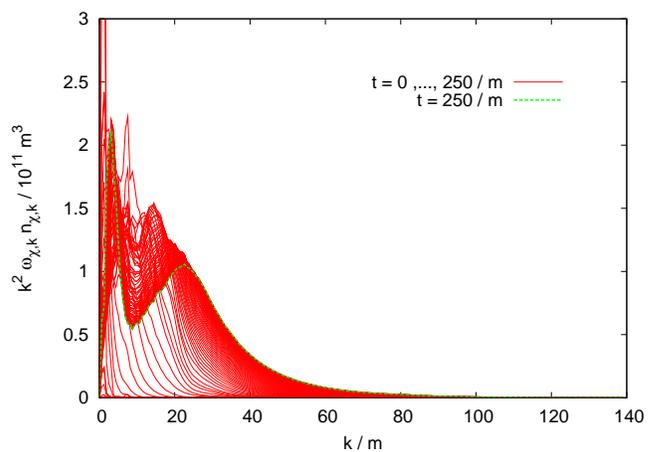}
\caption{Same as Figure \ref{fig:phispL}, but for the field $\chi$.}
\label{fig:chispL}
\end{figure}

In Figures~\ref{fig:phispL} and \ref{fig:chispL}, we plots the
distributions at different times separated by intervals $\delta t =
4\pi/m$. This allows us to visualize the growth of the distributions,
and to monitor the cascade of energy in the phase space $k$. When the
ultraviolet part of the distribution hits the highest momentum of the
simulation (defined by the grid size), the energy is artificially
reflected back to the IR modes and the simulation is no longer
reliable. Due to the scale chosen (natural rather than log scale),
only the distributions at times greater than about $100 /m$ can be
appreciated in the two figures~\ref{fig:phispL} and
\ref{fig:chispL}. Moreover, the double peak structure that can be
observed for $\chi$ at late times is due to the rescaling chosen.  The
plot for the occupation number, $n_k \, k^2$ (not shown) has a high
peak at $k \sim 4 m \,$, followed by a plateau up to $k \sim 30 \, m
\,$. The combination $k^2 \omega_k n_k$ shown has a peak at the
momenta corresponding to this plateau, showing that these momenta
dominate the energy density in the $\chi$ distribution. The saturated
spectra can be clearly seen in the green (thick) curves on these
plots.

\begin{figure}[h]
\leavevmode\epsfxsize=\columnwidth \epsfbox{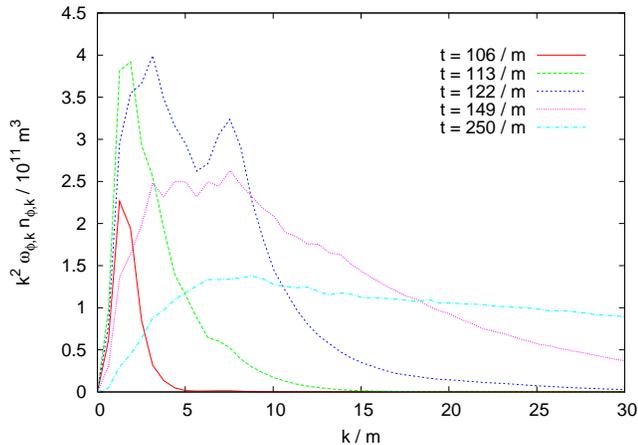}
\caption{Same combination as in Figure \ref{fig:phispL}, but zoomed in the IR region.}
\label{fig:phispS}
\end{figure}

\begin{figure}[h]
\leavevmode\epsfxsize=\columnwidth \epsfbox{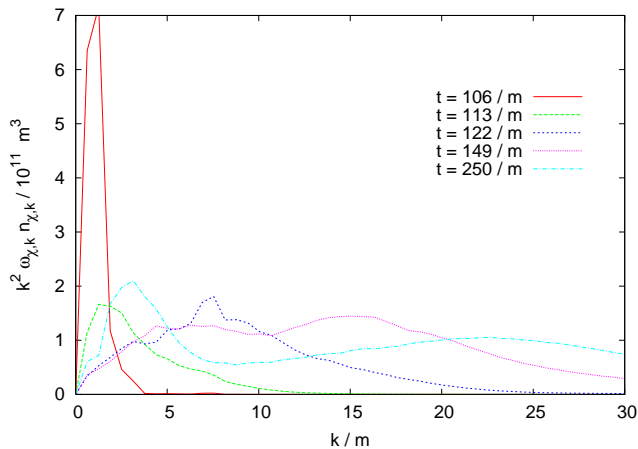}
\caption{Same as Figure \ref{fig:phispS}, but for the field $\chi$.}
\label{fig:chispS}
\end{figure}

The combination $k^2 \omega_k n_k$ is also given in
Figures~\ref{fig:phispS} and \ref{fig:chispS}, where we however show
only few times, and we focus on the IR part of the
distributions. These figures show the rapid broadening (towards the
UV) of the distributions in the violent rescattering stage, and the
tendency to saturation at later times.

\subsection{Fluctuations and Effective Masses}

The evolution of the scalar fields can be strongly affected by the
presence of dynamical effective masses for the two fields. A high mass
for some of the fields can disfavour or block some of the
interactions, with a consequent delay of thermalization. For instance,
it is well known that a condensate which is strongly coupled to some
field $\psi$ can have a very long lifetime, since the same coupling
which would allow the decay also produces a high effective mass for
$\psi$, which can prevent the decay from occuring. In the case we are
discussing, the effective masses of the fields acquire significant
(loop) contributions from the high variances produced at
preheating/rescattering, according to
Eqs.~(\ref{mass1})-(\ref{mass2}).

\begin{figure}[h]
\leavevmode\epsfxsize=\columnwidth \epsfbox{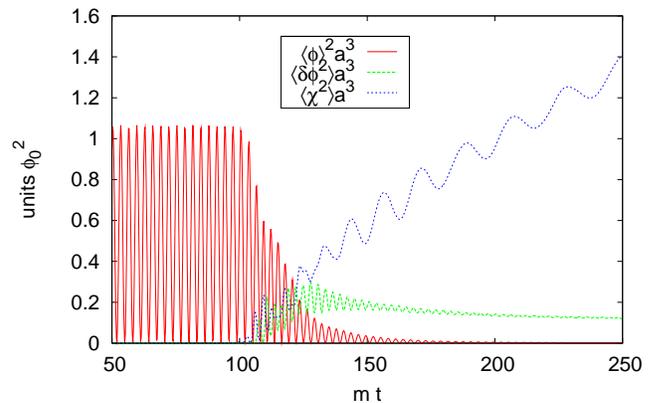}
\caption{Evolution of the background inflaton field $\langle\phi\rangle^2$ and of
the variances $\langle\delta \phi^2\rangle$ and $\langle\chi^2
\rangle$. Fields squared are rescaled by $a^3 \,$.}
\label{fig:fluc}
\end{figure}

\begin{figure}[h]
\leavevmode\epsfxsize=\columnwidth \epsfbox{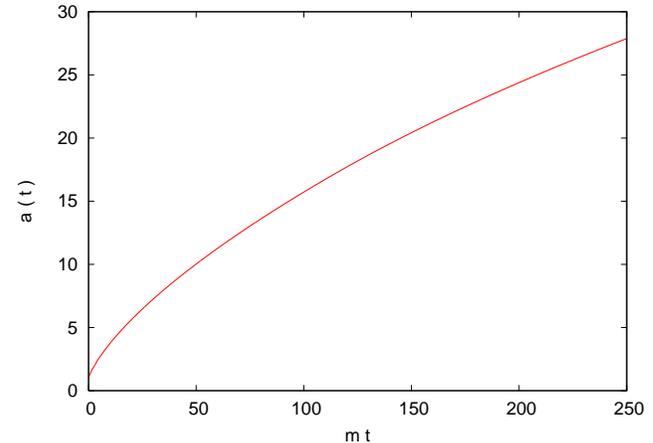}
\caption{Evolution of the scale factor $a(t)$.}
\label{fig:sf}
\end{figure}

Figure \ref{fig:fluc} shows the evolution of the background inflaton
oscillations and fluctuations of $\langle\delta \phi^2\rangle$ and
$\langle\delta \chi^2 \rangle$. All values are plotted in ``comoving
scales'', i.e. multiplied by $a^3$.  The corresponding physical
quantities can be obtained by the value of the scale factor, which is
shown in Figure \ref{fig:sf}.

\begin{figure}[h]
\leavevmode\epsfxsize=\columnwidth \epsfbox{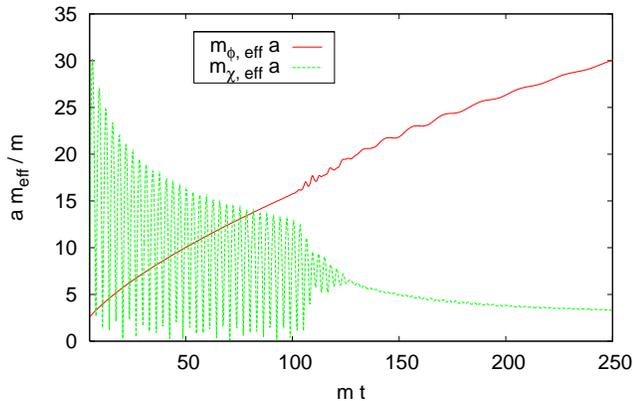}
\caption{Evolution of effective masses multiplied by scale factor $a$
(as it is relevant for the comoving frquency~(\ref{comom})).}
\label{fig:masses}
\end{figure}

Figure~\ref{fig:masses} shows the effective masses for $\phi$ and
$\chi \,$, multiplied by the scale factor $a$. This combination enters
in the comoving dispersion relations for the quanta of the two fields,
given by eq.~(\ref{comom}). Therefore, the product $a \, m_{\rm eff}$
is the correct quantity to compare with the comoving momentum $k \,$.

\begin{figure}[h]
\leavevmode\epsfxsize=\columnwidth \epsfbox{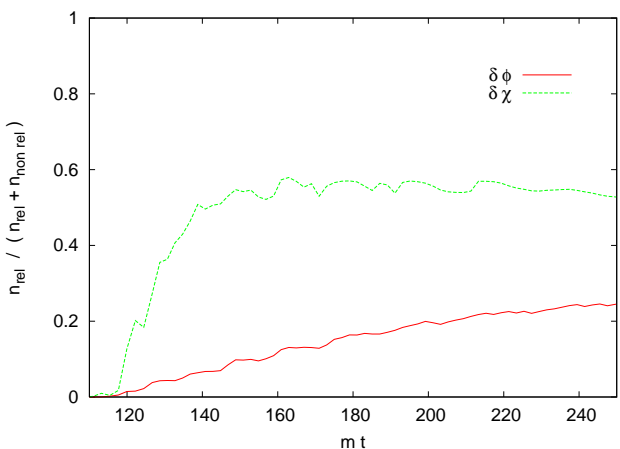}
\caption{Fraction of particles in the relativistic regime, $k > a \,
m_{\rm eff} \,$.}
\label{fig:frel}
\end{figure}

This comparison is shown in Figure~\ref{fig:frel}, where we show
the fraction of relativistic quanta (that is, with $k > a \, m_{\rm
eff}$) for the two fields. We see that the majority of the particles are
non-relativistic, so that the effective masses indeed play a
significant role in the evolution of the spectra. In particular, we
observe that during rescattering very few particles are relativistic; this
confirms the fact that the distributions are very peaked in the IR at
this stage. The fraction of relativistic $\chi$ particles rapidly increases
during the violent thermalization stage. The fraction of relativistic
$\delta \phi$ remains instead always smaller (confirming the fact that
$m_{\phi,{\rm eff}} > m_{\chi,{\rm eff}} \,$), and it slowly increases
during the thermalization stage, when the distributions evolve towards
the UV.

Up to $m t \sim 120 \,$, the coherent mode of the inflaton dominates
the effective mass for $\chi \,$. (The rapid variation of
$m_{\chi,{\rm eff}}$ due to the coherent oscillations of $\phi$ is the
cause of preheating.) At later times, the variance of $\phi$ gives
instead a greater contribution. In the effective mass for $\phi$ the
``bare'' term $m$ dominates up to $m t \sim 100 \,$, while the
variance of $\chi$ provides the dominant contribution at later
times. Eq.~(\ref{fluc3}) provides a useful expression for the
variance; during the turbulence stage, the occupation numbers vary
only adiabatically, so one may expect that the growth of $a \, m_{\rm
eff}$ results in a decrease of the variance, and vice-versa. This
relation is visible in the figures shown. The ``late time'' decrease
of $a \, m_{\chi,{\rm eff}}$ is accompanied by an increase of $\langle
\chi^2 \rangle$.  This in turns cause an increase of $a \,
m_{\phi,{\rm eff}}$, and, consequently, a decrease of $\langle \phi^2
\rangle$.

\section{Cosmological implementations of the results}\label{sec:implementations}

\subsection{Complete Decay of the Inflaton}

In the simple model discussed so far, with the interaction term
$\phi^2 \, \chi^2 \,$, the decay of the inflaton $\phi$ is not
complete. Indeed, such a term mediates scatterings between inflaton
quanta, rather than single particle decay. Once $\phi$ is diluted by
the expansion of the universe, the scatterings become inefficient, and
the number of inflaton quanta remain practically constant. This poses
a problem because in order to reach the stage of radiation domination
we need to have complete decay of the inflaton.

 The
simplest way to have a complete decay is to consider three legs
interactions. For instance, we can replace~(\ref{int}) by
\begin{equation}
{\cal L}_{\rm int} = - \frac{g^2}{2} \, \left( \phi + \sigma \right)^2
\chi^2
\end{equation}
where $\sigma$ is a mass dimension parameter, which breaks the $\phi
\leftrightarrow - \, \phi$ symmetry. For $\sigma \ll \phi \,$, the trilinear
term produced by this interaction is irrelevant and the inflaton decay occurs as described above.
However, when $\phi$ is decreased by the expansion of the universe,
the trilinear interaction becomes dominant, leading to a complete
inflaton decay.

We can avoid discussing the introduction of the scale $\sigma$ by
considering a Yukawa interaction of the inflaton with some fermions
$\psi \,$.  By itself, this interaction has interesting consequences
for preheating, as discussed e.g. in~\cite{fermi}. When both fermionic
and bosonic interactions are present, preheating into bosons is
typically much more efficient, since the one into fermions is reduced
by Pauli blocking.  However, once the interactions become
perturbative, the trilinear $\phi \rightarrow \psi \, \psi$ decay will
eventually dominate over the $\phi \phi \rightarrow \chi \chi$
scattering.

Three-legs interactions arise very naturally in SUSY theories. Indeed,
consider the simple superpotential
\begin{equation}\label{susy}
W=\frac{m}{\sqrt{8}}\phi^2+g^2\phi\chi^2 \ .
\end{equation}
The corresponding scalar field potential contains three-legs as well
as four-legs plus self-interaction terms
\begin{equation}\label{potl}
V=\frac{m^2}{2}\phi^2+ \sqrt{2} g^2 m \phi \chi^2 + 4 g^2 \phi^2 \chi^2+ g^4 \chi^4 \,
\end{equation}
(neglecting the imaginary parts of the fields). The four-legs
interaction dominates over the three-legs one as long as $\phi \gsim
\sigma=2g^2 m$. For reasonable values of $\sigma$, this happens during
the first stages of preheating. Eventually, the amplitude of $\phi$
decreases due to the expansion of the universe, and the trinlinear
interaction dominates, resulting in a complete decay of the massive
inflaton.

Hence, considering a supersymmetric theory automatically introduces
trilinear vertices among the scalars. It also allows for stronger
couplings between the inflaton and other fields, without spoiling the
flatness of the inflaton potential, as we have discussed in
Sec.~\ref{sec:model}.

\subsection{Relation between efoldings $N$ and wavelength of
perturbations}

The precise history of the expansion of the universe $a \left( t
\right)$ is important for connecting the physical wavelength $k / a \left(
t \right)$ of the cosmological perturbations we observe to the
number of efoldings $N$ at which the perturbation was generated (left
the horizon) during inflation.

According to the general lore \cite{LiddleBook,Liddle:1993fq},
 one has
\be
\label{efold}
N(k) = 62 - {\rm ln} \frac{k}{6.96\times 10^{-5} \, {\rm Mpc}^{-1}} +
\Delta ,
\ee
where $6.96\times 10^{-5} {\rm Mpc}^{-1}$ is the inverse size of the
present cosmological horizon and $\Delta$ is defined by the physics
after inflation:
\be
\Delta =-{\rm ln} \, \frac{10^{16} {\rm GeV}}{V_k^{1/4}}+
\frac{1}{4}{\rm ln} \frac{V_k}{V_{\rm end}}-
 \frac{1}{3} {\rm ln} \frac{V^{1/4}_{\rm end}}{\rho_{\rm reh}^{1/4}} \ ,
\label{DeltaDef}
\ee
where $\rho_{\rm reh}$ is the energy density at the end of reheating,
 $V_k$ is the value of the inflaton potential at the moment when the
 mode with the comoving wave number $k$ exits the horizon at inflation
 and finally $V_{\rm end}$ is the value at the end of inflation.

The last term in the right hand side of eq.~(\ref{DeltaDef}) is
related to  reheating, specifically that it  is not an instantaneous process. It
is
typically computed under the assumption that the universe is matter
dominated (by the coherent oscillations of the inflaton) all
throughout reheating, and that it (instantaneously) becomes radiation
dominated at the end of the reheating stage. As we have seen, this is
not what happens.

Indeed, a better definition of the last term in $\Delta$ is $
-\frac{1}{4} {\rm ln} \, \frac{a_{\rm rd}}{a_{\rm inf}}$ where $a_{\rm
  rd}$ is the scale factor at the moment after reheating at which the
EOS reaches its value at radiation dominance. As we have been
discussing, this moment can occur while the distributions of particles
are very far from thermal, shortly after the end of preheating. This
amounts to taking $a_{\rm rd} / a_{\rm inf} \sim 10 \mbox{---} 20 \,$,
rather than $a_{\rm rd} / a_{\rm inf} \gsim 10^8 \,$, where $a_{\rm
  RD}$ is the value of the scale factor at the end of reheating under
the common assumption that perturbative reheating occurs entirely
within a regime of matter domination, with $T_r \simeq 10^9 \, {\rm
  GeV}$. In this situation one has $H_ {\rm RD} \sim
\frac{T^2_r}{M_P}$ at the end of matter domination, and the
corresponding value of the scale factor $a_{\rm RD}$ can be estimated
as $\frac{a_{\rm RD}}{a_{\rm inf}} \sim \lmk \frac{H_{\rm inf}}{H_{\rm
    RD}}\rmk^{2/3} \sim \lmk \frac{m M_P}{T_r^2} \rmk^{2/3}$.

Hence, for a preheating scenario with a very wide range of values of
the coupling $g^2 \,$, the first term of~(\ref{DeltaDef}) can be fixed
to
\begin{equation}
- \frac{1}{4} {\rm ln} \, \frac{a_{\rm rd}}{a_{\rm inf}} \sim -
  (0.6\mbox{---}0.8) \,\,.
\label{finalrat}
\end{equation}
 This result holds for a wide class of inflationary models, and for
many different types of preheating.

\subsection{Physics before thermalization}

One of the most important parameters in physics of the early universe
is the highest temperature of the hot plasma after the inflaton field
decays and transfer its energy into radiation.\footnote{It is well
known that if you assume a slow, perturbative inflaton decay and a
quick thermalization of the decay products, a subdominant thermal bath
will be present while the inflaton is still decaying. That will have a
temperature muich higher than $T_r$, defined at the moment when the
thermal bath starts dominating.} This is traditionally called the
reheating temperature $T_r$. As we have seen in many previous papers
and above in this paper, ultimate thermal equilibrium is preceeded by
several stages. Let us distinguish four of them:

\begin{itemize}
\item preheating, with the duration $\delta t_1 \sim 100/m$,

\item a short violent stage at the end of preheating when
non-linearity and non-perturbative effects are dominant; chaos is
onset, erasing the details of the initial conditions, on a timescale
$\delta t_2 \sim 10/m$,

\item the stage of turbulence of classical fields, with the saturated
spectrum cascading both towards UV and IR modes with duration $\delta
t_3$, longer than the two previous stages,

\item the last stage of proper QFT thermalization, with particle
fusion/offshell processes; it is characterized by the timescale
$\delta t_4$, much longer than the previous ones.
\end{itemize}

Traditionally the beginning of the thermal history is related
to the moment $\delta t_4$, which may be very long; as a result of
redshift, the reheating temperature can be rather low.

Let us reconsider this attitude. While the question of when the
ultimate thermal equilibrium will be established and to which value of
$T_r$ it corresponds is still very interesting, we will argue that
important physics which constrains the model of inflation and
interactions takes place long before full thermal equilibrium is
completed.

Indeed, Figures~\ref{fig:phinkwk}--\ref{fig:phispL} show that the
modes at physically interesting scales, up to $k \sim$ tens of $m \sim
10^{14}$ GeV, are quickly excited towards a saturated distribution.
This raises several subtle and interesting questions. Usually, the
relativistic particles embedded in the thermal bath are brought into
thermal equilibrium after the relaxation time $1/\sigma n$, where
$\sigma$ is the cross section of the processes relevant for
thermalization, and $n=\int d^3k n_k$ is the number density of
particles. This estimate works in the diluted gas
approximation. Indeed, consider the vertex $\phi \, \chi^2$, leading
to inflaton decay into quanta of $\chi$. The particle number density
enters in the collisional integral in the combination
\begin{equation}
n_{p1}^\phi \left( 1 + n_{p2}^\chi \right) \left( 1 + n_{p3}^\chi \right) -
n _{p2}^\chi \, n_{p3}^\chi \left( 1 + n_{p1}^\phi \right)
\label{collfactor}
\end{equation}
($p_1$ is the momentum of the inflaton quantum, while $p_{2,3}$ are
the momenta of the two quanta of $\chi$ entering in the process). The
above estimate for the thermalization timescale assumes that the
inverse process $\chi \chi \rightarrow \phi$ (described by the second
term in~(\ref{collfactor})) is irrelevant, and that all the occupation
numbers are much smaller than one, so that $1+ n_p \simeq 1 \,$. In
particular, this last assumption is not valid until the very end of
the thermalization stage.

Therefore, for large $n_k$ we expect that particles are dragged into
effective thermal equilibrium faster, due to the stimulated
interaction. We have observed this fact also in previous numerical
lattice simulations \cite{FK}.

Another relevant effect, already discussed above, is that large
fluctuations of fields can contribute to the effective masses of
particles, which can become effectively heavy. This leads to the
blocking of some processes which may lead to a faster thermalization,
until fields fluctuations are diluted by the expansion.

Finally, a very relevant question is the one of particle production
from the nonthermal but highly excited distributions. In particular,
one should verify that dangerous relics are not overproduced at this
stage. To see this, let us discuss gravitino production from the decay
products of the inflaton. In models where supersymmetry is broken
gravitationally, only a very small number of gravitinos can be
tolerated, $Y_{3/2} = n_{3/2} / s \lsim 10^{-14}$ (the exact value
being dependent on the gravitino mass) \cite{grath}\footnote{A more
stringent limit applies if the decay has a significant hadronic
branching ratio \cite{moroi}.}.  The typical approach is to compute the
production of gravitinos only after the thermal bath has formed. The
dominant processes are $2 \rightarrow 2$ scatterings with only one
gravitino as outgoing particle (and, hence, only one gravitationally
suppressed vertex, while the other vertex is typically a gauge
interaction). The rate of gravitino production follows from the
Boltzman equation
\begin{equation}
\frac{d N_{3/2}}{d t} + 3 \, H \, N_{3/2} = \langle \sigma \, \vert v
\vert \rangle \, N_T \, N_T \,\,,
\end{equation}
where $N_T \sim T^3$ is the abundance of the (MSSM) degrees of freedom
in the thermal bath, and where the average cross section is of the
order $\langle \sigma \vert v \vert \rangle \sim \alpha / M_p^2 \,$,
where $\alpha$ is a gauge structure constant. The right hand side
decreases very rapidly with time, and the production at the highest
possible temperature (that is, at $T_r$) dominates. Hence, only the
production in the first Hubble time is relevant, and the final result
can be estimated as
\begin{equation}
N_{3/2} \sim \, \frac{\alpha \, T_r^6}{M_p^2} \times H_r^{-1} \,\,,
\end{equation}
where $H_r \sim 10 \, T_r^2 / M_p$ is the Hubble parameter at $T = T_r
\,$. This leads to the gravitino abundance 
\begin{equation}
Y_{3/2} \sim 10^{-4} \, \frac{T_r}{M_p} \,\,,
\label{grat}
\end{equation}
and to the well known bound $T_r \lsim 10^9 \, {\rm GeV}$ on the
reheating temperature.

This standard computation of gravitinos generated from the thermal
bath neglects all the production which may have taken place during the
thermalization stage. The underlying idea is that, during reheating,
most of the energy is in the inflaton condensate, where the quanta
(loosely speaking) do not have momentum, and hence cannot scatter to
produce gravitinos. While this may be true for a slow (perturbative)
inflaton decay, this is certainly not the case for
preheating/rescattering, where the distributions of the decay products
form right after the end of inflation. These distibutions may be
responsible for significant gravitino production, even before complete
thermalization has taken place~\cite{BDPS}.

The computation is now more difficult than in the thermal
case~\cite{grath}. First of all, unlike the thermal computation, the
result is expected to be model dependent. Second, turbulent nonlinear
processes are dominant right after preheating, so that it is possible
that processes including more vertices give a sizeable contribution to
the amount of gravitinos produced. However, we can obtain a
conservative estimate by considering only $2 \rightarrow 2$ tree level
processes also in this case. For definiteness, let us take the model
(\ref{pot}), (\ref{int}), with $g^2 = 2.5 \times 10^{-7} \,$ (the value
that we have studied above), and assume that the $\chi$ field produced
at rescattering has a trilinear vertex with its femionic superpartner
${\tilde \chi}$ and another fermion ${\tilde z}$.  For instance, this
could be a gaugino, if $\chi$ has gauge interactions. However, let us
consider here a generic vertex with coupling constant $h\,$ and the
interaction $\chi \, \chi \rightarrow \psi_{3/2} \, {\tilde z}$,
mediated by the fermionic partner of $\chi$. (As a technical point, we
consider production of transversal gravitinos, since the nature of the
longitudinal component changes with time).

We are interested in evaluating the production at the time $t_* \sim
120 / m \,$, when the (physical) number density of quanta of $\chi$ is
maximal. We first discuss whether the process we are considering is
kinematically allowed. At the time $t_* \,$, the non-gravitational
vertex provides an effective mass for fermions of the order (see
Fig.~\ref{fig:fluc}).
\begin{equation}
m_{\tilde z} \left( t_* \right) \sim h \, \sqrt{\langle \chi^2 \left(
t_* \right) \rangle} \sim 5 \cdot 10^{2} \, h \, m
\end{equation}
The distributions of $\chi$ obtained from the lattice have a typical
comoving momenum of the order of $10 \, m \,$, which, at the time $t_*
\,$, corresponds to the physical momentum $p_* \sim 0.5 \, m
\,$. Hence, if $h < 10^{-3} \,$, most of the quanta of $\chi$ will be
able to produce gravitinos through the process we are
considering. For higher values (for example, if the second vertex is a
gauge interaction), only the few quanta of $\chi$ with very high
momenta can contribute to the production, so that this process is
significantly suppressed. However, as the universe expands,
$\sqrt{\langle \chi^2 \rangle}$ decreases as $a^{-3/2} \,$,
while the physical momenta $p \propto a^{-1}$ decrease
less. Therefore, more and more quanta will have a sufficiently high
momentum to produce a $\psi_{3/2} \mbox{--} {\tilde z}$
pair.\footnote{This scaling argument does not include the effect of
the thermalization, which is also increasing the $p / m_{\tilde z}$
ratio. Indeed, since thermalization proceeds through particle fusion it
has the effect of both increasing the typical momenta of the
distributions and of decreasing the effective masses.} For
simplicity, $h \lsim 10^{-3}$ is assumed in the following discussion;
we do not expect that the following conclusions will significantly
change if a stronger coupling is considered.

By proceeding as in the thermal case, we can estimate the number of
gravitinos produced in the first Hubble time after $t_*\,$. (Again, the
numerical values are taken from the lattice results; $N_\chi \left(
t_* \right)^2$ denotes the physical number density of the $\chi$
quanta at the time $t_*\,$.)
\begin{equation}
N_{3/2} \left( t_* \right) \sim \frac{h}{M_p^2} \, N_\chi \left( t_*
\right)^2 \, H \left( t_* \right)^{-1} \sim 10^7 \, h \, m^3 \,\,.
\label{n321}
\end{equation}
For sufficiently high $h \,$, this value is actually incorrect, since
it exceeds Pauli blocking. Assuming that the whole Pauli sphere up to
physical momentum $p_*$ is filled gives
\begin{equation}
N_{3/2} \left( t_* \right) \sim 0.1 \, m^3.
\label{n322}
\end{equation}
The number density of produced gravitinos is then the smaller between
eqs.~(\ref{n321}) and~(\ref{n322}), according to the value of $h
\,$. In the following, we assume $h > 10^{-8} \,$, so that Pauli
blocking is reached and the number density of gravitinos is given
by eq.~(\ref{n322}).

The gravitino abundance is obtained by dividing the gravitino number
density by the entropy. At the time $t_* \,$, the distributions of the
decay products are far from thermal (being concentrated in the IR), so
that their entropy is smaller than that of a thermal bath with the
same energy. The entropy density can be computed from the lattice
simulations to be approximately $s=10^4 \, m^3$ at this
stage~\cite{marco}. However, the (comoving) entropy is expected to
increase due to thermalization.  If we convert the actual particle
distribution instanteneously into the thermal distribution, the
entropy would reach the maximal value $s_{max} \sim 10^7\, m^3$
allowed by energy conservation.  The difference between the actual
entropy density and $s_{max}$ gives rise to the dilution of the
gravitino abundance during thermalization. (Further sources of
dilution require extensions of the minimal model.)

Therefore, the conservative estimate for the gravitino abundance
is obtained by dividing the gravitino number density by
\begin{equation}
\rho \left( t_* \right)^{3/4} \sim 10^7 \, m^3 \,\,.
\end{equation}
where $\rho$ is the (physical) background energy density at the time
$t_* \,$. Hence, the ratio
\begin{equation}
Y_{3/2} \left( t_* \right) = \frac{N_{3/2} \left( t_*
\right)}{\rho^{3/4} \left( t_* \right)} \sim 10^{-8}
\label{granont}
\end{equation}
is about $6$ orders of magnitude higher than the allowed limit. 

This is a serious concern, which has to be addressed in the model
building of inflation. It can be resolved by introduction of extra 
radical assumptions.
For instance, this gravitino abundance can be
decreased by an entropy dilution.  This happens if a massive species
$\psi$ is produced at reheating, and it dominates the energy density
of the universe for some time. When it decays, it leads to a
significant amount of entropy, which dilutes the particle pieces
produced directly at the inflaton decay.  Concretely, the entropy
dilution leads to the decrease of the gravitino abundance $Y_{3/2}
\sim N_{3/2} / \rho^{3/4}$. Supposing that the universe is matter
dominated between the times $t_1$ and $t_2 \,$, the ratio $N_{3/2} /
\rho^{3/4}$ is decreasing as $\left( a_1 / a_2 \right)^{3/4} \;$
($a_i$ being the scale factor at the time $t_i$). This entropy
dilution could also occur through a secondary stage of
inflation \cite{inflationafterpreheating}.

In the specific toy model we have considered, a decrease of six orders
of magnitude does not seem to be realistic, so that the production of
gravitinos in this model appears to be a serious concern. However, it
is important to remember that the result~(\ref{granont}) is very
sensitive to the specific model considered, and that a case by case
calculation is necessary.

\subsection{Modulated Fluctuations from Preheating}

One of the motivations for studying the EOS after preheating is
related to the theory of modulated cosmological fluctuations.  We
briefly review the idea as it was presented in~\cite{mod, mod1}, and
discuss the implications of our findings on the evolution of the EOS
for this mechanism. We leave the explicit calculations for subsequent
publication.
 
Suppose that the coupling $g^2$ depends on some modulus field $z$, (as is typical in string theory)
\begin{equation}
g^2=g^2(z) \ , \,\,\,  \delta g^2=\frac{d g^2}{d z} \, \delta z \ .
\end{equation}
During inflation the modulus field $z$ can be light, with a mass
smaller than the Hubble parameter during inflation $H \sim 10^{13}$
Gev. Then, large scale fluctuations $\delta z$ with an almost scale
free spectrum are inevitably generated from inflation.  The wavelength
of fluctuations exceeds the size of the causal patch $H^{-1}$. As we
have discussed above, the EOS varies very quickly already at the
beginning of the thermalization stage. The variation is a non
monotonic function of the coupling $g\,$, see for instance
Figs.~\ref{fig:w1} and~\ref{fig:tran}. Due to the large scale spatial
fluctuations of the coupling $\delta g^2$, the change in the EOS
proceeds with a slight time shift in different Hubble patches. This
results in the generation of (almost) scale free scalar metric
cosmological perturbations.
 
To perform accurate calculations of the amplitude of modulated
fluctuations, one needs to know the exact character of the transition
from inflaton field to radiation.  In the simplified picture of
perturbative reheating, when the inflaton energy decays exponentially
as $\epsilon_{\phi} \sim e^{-\Gamma t}$, calculations of modulated
fluctuations~\cite{mod1} are based on linearization of the inflaton
decay rate $\Gamma$, which is linearly dependent on the coupling $g^2$.
However, as we have remarked, the decay of the inflaton field is
typically more complicated, and the transition towards RD occurs very
quickly (well before thermalization completes). In addition, we stress
that the time of the transition is a non monotonic function of
$g^2$. Finally, although we focused only on the direct coupling of the
inflaton to matter, the details of thermalization (and of the
evolution of the EOS) depend on the other interactions among the decay
products; the strength of these interactions are likely to be
dependent on some modulus field as well, and they can also modulate
cosmological perturbations. All these features have to be taken into
account for precision computations of modulated fluctuations.

\section{Summary}\label{sec:summary}

In this paper we studied the out-of-equilibrium nonlinear dynamics of
the fields during and after preheating. This continues earlier works
in \cite{FK,MT04,Berges:2004}.  We use lattice numerical simulations
of fully non-linear dynamics of interacting classical fields. In this
investigation, we computed the evolution of the equation of state
during preheating and the early thermalization stage.  Immediately
after inflation, the EOS is the one corresponding to inflaton
domination.  For a massive inflaton, we have matter domination,
characterized by $p=0$.  After a very short time, the inflaton
transfers its energy into inhomogeneous modes. At the first stage
(preheating) inhomogeneous field fluctuations are copiously produced
in the regime of parametric resonance. Then the modes rescatter to
redistribute the energy by cascading in the phase space.  This process
of classical wave turbulence can be rather long. This leads to a final
stage, where the occupation numbers are small and the classical
approximation breaks down. Thermalization finishes when the quantum
fields reach the ultimate thermal equilibrium. If all the
participating fields are light, the final thermalized state
corresponds to an EOS of radiation, with $p=\frac{1}{3}\epsilon$.

The transition between the matter and radiation dominated EOS is hence
typically supposed to occur on a (relatively) very long timescale.
What we found in our numerical simulations is very different from this
naive expectation. In fact, the EOS jumps from matter domination to
(almost) radiation domination immediately after the first stage of
preheating, i.e. in a couple dozen inflaton oscillations
($10^{-37}$ sec each), long before thermal equilibrium has been
established.  In other words, the macroscopic EOS is close to the one
of radiation domination, while the microscopic state is far from
thermal. This result is similar to the conclusion of
\cite{Berges:2004} regarding $O(N)$ theory in flat spacetime.

The sharp change of the EOS we observed can be very conveniently
related to the moment of energy transfer from the inflaton to the
inhomogeneous radiation.  As we mentioned, it is a very fast process,
as is well-known from the theory of preheating, which takes place in
only about $10^{-35}$ sec. Moreover, the time at which the transition
occurs depends only weakly on the coupling $g^2$ between the inflaton
and the other fields. The exact dependance is non-monotonic, and the
transition time oscillates around $\sim 10^{-35}$ sec (as set by
preheating) for a very wide range of couplings. This result is
drastically different from perturbative inflaton decay, where the
transition timing is proportional to the coupling.

This very quick evolution has several consequences, and leads to a
number of issues which deserve further investigation. We stress four
of them.

i) The first is related to the generation of modulated cosmological
fluctuations from preheating. It is conceivable~\cite{mod,mod1,BKU} that the
strength of the inflaton interactions with its decay products is a
function of some moduli fields as expected in string
theory. Fluctuations of the light moduli (with mass smaller than
$10^{13}$ GeV) are inevitably generated during inflation. This results
in spatial variations of the couplings at large scales, well beyond
the size of the causal (Hubble) patch at the end of inflation. Hence,
the transition from inflation to radiation domination occurs at
slightly diffferent times in different Hubble patches, giving rise to
scalar metric fluctuations.

To calculate metric fluctuations from modulated perturbations, one
needs to know the exact evolution of the EOS, from the matter
domination (inflaton) to the radiation domination (light decay
products) stage.  We found that the transition of the EOS occurs
sharply in a step-like manner. Therefore one can introduce a time-like
(but spatially rippled) hypersurface which divides two cosmological
regimes, with $p=0$ and $p=\frac{1}{3} \epsilon$.  Metric fluctuations
can be calculated using GR matching conditions across this
hypersurface, similar to how it was done to calculate modulated
fluctuations in hybrid inflation \cite{BKU}.

ii) The second application of the EOS history is related to the
formula which links the number of efoldings during inflation $N$ and
the present day physical wavelengths of fluctuations in terms of the
logs of their momenta $\log k$. Indeed, the cosmological evolution of
the scale factor $a \left( t \right)$ is ultimately defined by the
EOS. Refining the $N-\log k$ relation is mandatory to put precise
observational constraints on inflationary models. In this paper we
notice that a (practically) immediate transition from inflation to
radiation domination is a common feature of inflationary models with
preheating, either parametric resonance or tachyonic
ones. Equation~(\ref{DeltaDef}), with our result~(\ref{finalrat}),
refines the $N-\log k$ relation.

iii) The next note is related to the model building of inflaton
interactions. In the majority of papers discussing inflaton decay,
interactions with bosons are considered in the form of four-legs
vertices of the type $g^2 \phi^2 \chi^2$. These interactions are the
dominant ones at the early stages of preheating, when the amplitude of
the inflaton is large. (We do not consider here nonrenormalizable
interactions, which -- if present -- could be more important.)
However, they do not lead to a complete inflaton decay. A complete
decay requires vertices with only one inflaton, as the Yukawa
interaction $\phi \, {\bar \psi} \, \psi$ wih fermions, or the
three-legs bosonic interaction $\sigma \, \phi \, \chi^2$ (where
$\sigma$ is a mass scale). As we have argued, these trilinear
interactions are very natural in supersymmetric models. Supersymmetry
has the additional advantage of protecting the flatness of the
inflaton potential from too large radiative corrections, which arise
from the coupling of the inflaton to the light degrees of freedom.

iv) Finally, the early evolution of the light degrees of freedom from
the decay of the inflaton can have very important consequences for the
generation of particles far from thermal equilibrium.  A particularly
relevant application is the possible (over)production of gravitinos
from the decay products of the inflaton. Although this is a
perturbative generation, it can be very effective, since it occurs at
very early times. Traditionally, the limit $T_r \lsim 10^9 \, {\rm
GeV}$ is due to gravitinos produced only after the $\chi$ quanta have
thermalized.  However, a nonthermal distribution can also be
responsible for gravitino overproduction~\cite{BDPS}. These inflaton
decay products arise very rapidly, with an energy density much higher
than $\left( 10^9 \, {\rm GeV} \right)^4$; although their initial
distribution is far from thermal, we have seen that they can be
responsible for a significant gravitino production, which in some
cases exceeds by several orders magnitude the allowed bound. Clearly,
this is a less universal bound than the standard one, since it depends
on the details of how the thermal bath is produced. (For instance, the
gravitino abundance can be diluted by entropy release during
thermalization.) In addition, the gravitino problem is avoided
altogether if the gravitino mass is significantly smaller (as in gauge
mediated supersymmetry breaking) or larger (as in the simplest
versions of the KKLT model~\cite{KL}) than the electroweak
scale. Still, this possible production must be addressed in many
inflationary models.

\vspace{1cm}

{\bf Acknowledgements}. We thank J. Berges, J.F. Dufaux, A. Linde,
A. Mazumdar, and I. Tkachev for useful discussions. GF and MP thank
CITA for its warm hospitality during different stages of this work.
LK was supported by CIAR and NSERC. MP was supported in part by the
DOE grant DE-FG02-94ER40823.

\appendix



\begin{thebibliography}{99}


\bibitem{DL}
A.~D.~Dolgov and A.~D.~Linde,
Phys.\ Lett.\ B {\bf 116}, 329 (1982);
%
L.~F.~Abbott, E.~Fahri, and M~Wise, Phys.\ Rev.\ Lett. {\bf 117B}, 29 (1982).


\bibitem{therma}
R.~Allahverdi,
Phys.\ Rev.\ D {\bf 62}, 063509 (2000)
[arXiv:hep-ph/0004035];
%
S.~Davidson and S.~Sarkar,
JHEP {\bf 0011}, 012 (2000)
[arXiv:hep-ph/0009078];
%
R.~Allahverdi and M.~Drees,
Phys.\ Rev.\ D {\bf 66}, 063513 (2002)
[arXiv:hep-ph/0205246].


\bibitem{GKR}
G.~F.~Giudice, E.~W.~Kolb and A.~Riotto,
Phys.\ Rev.\ D {\bf 64}, 023508 (2001)
[arXiv:hep-ph/0005123].

\bibitem{KLS94} L.~Kofman, A.~D.~Linde and A.~A.~Starobinsky,
Phys.\ Rev.\ Lett.\  {\bf 73}, 3195 (1994)
[arXiv:hep-th/9405187].

\bibitem{TB} J.~Traschen and R.~Brandenberger, Phy.\ Rev.\ D {\bf 42}, 2491 (1990).

\bibitem{KLS97} L.~Kofman, A.D.~Linde and A.A.~Starobinsky, Phys.\
Rev. {\bf D56} 3258 (1997).

\bibitem{tach}
G.~N.~Felder, J.~Garcia-Bellido, P.~B.~Greene, L.~Kofman, A.~D.~Linde and I.~Tkachev,
Phys.\ Rev.\ Lett.\ {\bf 87}, 011601 (2001) [arXiv:hep-ph/0012142];
G.~N.~Felder, L.~Kofman and A.~D.~Linde,
Phys.\ Rev.\ D {\bf 64}, 123517 (2001) [arXiv:hep-th/0106179].


\bibitem{mod}
L.~Kofman,
arXiv:astro-ph/0303614.

\bibitem{mod1}
G.~Dvali, A.~Gruzinov and M.~Zaldarriaga,
Phys.\ Rev.\ D {\bf 69}, 023505 (2004)
[arXiv:astro-ph/0303591].

\bibitem{BKU} F.~Bernardeau, L.~Kofman and J.~P.~Uzan,
arXiv:astro-ph/0403315.


\bibitem{KT}
S.~Y.~Khlebnikov and I.~I.~Tkachev,
Phys.\ Rev.\ Lett.\  {\bf 77}, 219 (1996)
[arXiv:hep-ph/9603378].

\bibitem{FK}
G.~N.~Felder and L.~Kofman,
Phys.\ Rev.\ D {\bf 63}, 103503 (2001)
[arXiv:hep-ph/0011160].

\bibitem{MT04} R.~Micha and I.~I.~Tkachev, Phys.\ Rev. {\bf D70},
043538 (2004).  [arXiv:hep-ph/0403101]; R.~Micha and I.~I.~Tkachev, Phys.\ Rev.\ Lett. {\bf 90} 121301 (2003).

\bibitem{FT}
G.~N.~Felder and I.~Tkachev,
arXiv:hep-ph/0011159.

\bibitem{Boyanovsky:1997cr}
  D.~Boyanovsky, D.~Cormier, H.~J.~de Vega, R.~Holman, A.~Singh and M.~Srednicki,
  Phys.\ Rev.\ D {\bf 56}, 1939 (1997)
  [arXiv:hep-ph/9703327].

\bibitem{J}
J.~Berges and J.~Serreau,
arXiv:hep-ph/0302210;
J.~Berges and J.~Serreau,
arXiv:hep-ph/0410330.

\bibitem{Berges:2004}
J.~Berges, S.~Borsanyi and C.~Wetterich,
Phys.\ Rev.\ Lett.\  {\bf 93}, 142002 (2004)
[arXiv:hep-ph/0403234].



\bibitem{grapreh}
R.~Kallosh, L.~Kofman, A.~D.~Linde and A.~Van Proeyen,
Phys.\ Rev.\ D {\bf 61} (2000) 103503
[arXiv:hep-th/9907124],
%
Class.\ Quant.\ Grav.\  {\bf 17} (2000) 4269
[arXiv:hep-th/0006179];
%
G.~F.~Giudice, A.~Riotto, and I.~Tkachev
JHEP {\bf 9908} (1999) 009
[arXiv:hep-ph/9907510],
%
JHEP {\bf 9911} (1999) 036
[arXiv:hep-ph/9911302];
%
H.~P.~Nilles, M.~Peloso and L.~Sorbo,
Phys.\ Rev.\ Lett.\  {\bf 87} (2001) 051302
[arXiv:hep-ph/0102264],
%
JHEP {\bf 0104} (2001) 004
[arXiv:hep-th/0103202].

\bibitem{grath}
J. Ellis, A. Linde and D. Nanopoulos, Phys. Lett. {\bf 118B}, 59
(1982); L.M. Krauss, Nucl. Phys. {\bf B227}, 556 (1983); D.
Nanopoulos, K. Olive and M. Srednicki, Phys. Lett. {\bf 127B}, 30
(1983); M. Yu. Khlopov and A. Linde, Phys. Lett. {\bf 138B}, 265
(1984); J. Ellis, J. Kim and D. Nanopoulos, Phys. Lett. {\bf
145B}, 181 (1984);   J. Ellis, G.B. Gelmini, J.L. Lopez, D.V.
Nanopoulos, and S. Sarkar, Nucl. Phys. {\bf B373}, 399 (1992);
M.~Kawasaki and T.~Moroi, Progr. Theor. Phys. {\bf 93}, {879}
({1995}).

\bibitem{new}
J.F.~Dufaux, G.~Felder, L.~Kofman, M.~Peloso, D.~Podolsky, in preparation.

\bibitem{KL87}
L.~A.~Kofman and A.~D.~Linde,
Nucl.\ Phys.\ B {\bf 282}, 555 (1987).


\bibitem{ST}
D.~V.~Semikoz and I.I. Tkachev, Phys.\ Rev.\ Lett. {\bf 74} 3093 (1995), [arXiv:hep-ph/9409202].

\bibitem{fermi}
J.~Baacke, K.~Heitmann and C.~Patzold,
Phys.\ Rev.\ D {\bf 58}, 125013 (1998)
[arXiv:hep-ph/9806205];
P.~B.~Greene and L.~Kofman,
Phys.\ Lett.\ B {\bf 448}, 6 (1999)
[arXiv:hep-ph/9807339];
G.~F.~Giudice, M.~Peloso, A.~Riotto and I.~Tkachev,
JHEP {\bf 9908}, 014 (1999)
[arXiv:hep-ph/9905242].
P.~B.~Greene and L.~Kofman,
Phys.\ Rev.\ D {\bf 62}, 123516 (2000)
[arXiv:hep-ph/0003018].
M.~Peloso and L.~Sorbo,
JHEP {\bf 0005}, 016 (2000)
[arXiv:hep-ph/0003045].


\bibitem{LiddleBook}
A.~R.~Liddle and D.~H.~Lyth, {\it Cosmological Inflation and Large Scale structure},
Cambridge University Press, Cambridge, 2000.

\bibitem{Liddle:1993fq}
A.~R.~Liddle and D.~H.~Lyth,
Phys.\ Rept.\  {\bf 231}, 1 (1993)
[arXiv:astro-ph/9303019].





\bibitem{moroi} M.~Kawasaki, K.~Kohri, and T.~Moroi,
[arXiv:astro-ph/0402490].

\bibitem{BDPS}
L.~Boubekeur, S.~Davidson, M.~Peloso and L.~Sorbo,
Phys.\ Rev.\ D {\bf 67}, 043515 (2003)
[arXiv:hep-ph/0209256].

\bibitem{marco} For the entropy density we used the formula $s=\int
d^3 k \left[\left(n_k+1\right)\rm{ln}\left(n_k+1\right)-n_k \rm{ln}
\left(n_k\right)\right]$. See for example R.~H.~Brandenberger, T.~Prokopec and V.~Mukhanov,
  Phys.\ Rev.\ D {\bf 48}, 2443 (1993)
  [arXiv:gr-qc/9208009].

\bibitem{inflationafterpreheating}
D.~H.~Lyth and E.~D.~Stewart,
  Phys.\ Rev.\ Lett.\  {\bf 75}, 201 (1995)
  [arXiv:hep-ph/9502417];
L.~Kofman, A.~D.~Linde and A.~A.~Starobinsky,
  Phys.\ Rev.\ Lett.\  {\bf 76}, 1011 (1996)
  [arXiv:hep-th/9510119];
  G.~N.~Felder, L.~Kofman, A.~D.~Linde and I.~Tkachev,
  JHEP {\bf 0008}, 010 (2000)
  [arXiv:hep-ph/0004024].


\bibitem{KL}
R.~Kallosh and A.~Linde,
JHEP {\bf 0412}, 004 (2004)
[arXiv:hep-th/0411011].


\end{thebibliography}
\end{document}